\documentclass{aastex6}
\usepackage{multirow}
\begin{document}

\title{The Energy Sources of Double-peaked superluminous supernova PS1-12cil and luminous supernova SN 2012aa}
\author{Long Li\altaffilmark{1,2,3}, Shan-Qin Wang\altaffilmark{1}, Liang-Duan Liu\altaffilmark{4},
Xiang-Gao Wang\altaffilmark{1}, En-Wei Liang\altaffilmark{1}, and Zi-Gao Dai\altaffilmark{2,3}}

\begin{abstract}

In this paper, we present the study for the energy reservoir powering the light curves (LCs) of
PS1-12cil and SN 2012aa which are superluminous and luminous supernovae (SNe), respectively.
The multi-band and bolometric LCs of these two SNe show unusual secondary bumps after the main
peaks. The two-peaked LCs cannot be explained by any simple energy-source models (e.g., the
$^{56}$Ni cascade decay model, the magnetar spin-down model, and the ejecta-circumstellar medium
interaction model). Therefore, we employ the $^{56}$Ni plus ejecta-circumstellar medium (CSM)
interaction (CSI) model, the magnetar plus CSI model, and the double CSI model to fit
their bolometric LCs, and find that both these two SNe can be explained by
{the double CSI model} and the magnetar plus CSI model.
Based on the modeling, we calculate the the time when the shells were expelled
by the progenitors{: provided that they were powered by double ejecta-shell CSI, the
inner and outer shells} might be expelled {$\sim 0.2-3.6$ and $\sim 2-25$ years
before the explosions of the SNe, respectively;} {the shells were expelled}
$\sim 2-20$ years before the explosions of the SNe {if they were powered by
magnetars plus CSI}.
\end{abstract}


\affil{\altaffilmark{1}Guangxi Key Laboratory for Relativistic Astrophysics,
School of Physical Science and Technology, Guangxi University, Nanning 530004,
China; shanqinwang@gxu.edu.cn; wangxg@gxu.edu.cn}
\affil{\altaffilmark{2}School of Astronomy and Space Science, Nanjing
University, Nanjing 210093, China; dzg@nju.edu.cn}
\affil{\altaffilmark{3}Key Laboratory of Modern Astronomy and Astrophysics
(Nanjing University), Ministry of Education, China}
\affil{\altaffilmark{4}Department of Astronomy, Beijing Normal University,
Beijing 100875, China}

\section{Introduction}

Supernovae (SNe) are energetic explosions emitting luminous ultraviolet (UV), optical,
and near infrared (NIR) radiation. The peak luminosities of most of SNe
are $\sim 10^{42}-10^{43}$ erg s$^{-1}$.
According to the spectra around the light-curves (LCs) peaks, SNe can be
divided into types I which are hydrogen-deficient and II which are
hydrogen-rich \citep{Minkowski41}. Type I SNe can be divided into
types Ia, Ib, and Ic, while most of type II SNe can be divided into
types II-P, II-L, IIn, and IIb. \citep{Filippenko97}.
The spectral types directly reflects the properties of the progenitors of SNe.
It has long been believed that type Ia SNe result from the explosions of white
drawfs \citep{Hoyle60}, while the other types of SNe come from collapse of
massive stars \citep{Baade34,Janka16}.

SNe Ic have attracted more and more attention due partly to the fact that
a fraction of them have been found to be associated with long gamma-ray bursts (GRBs)
(e.g., \citealt{Galama98,Hjorth03}, see \citealt{Woosley06,Cano17} for reviews).
On the other hand, over 100 superluminous SNe (SLSNe) 
whose peak luminosities are $\sim$ 100 times brighter than that of normal SNe
have been discovered and studied in the past two decades\citep{GalYam12,Inserra19};
meanwhile, dozens of luminous optical transients (including luminous SNe)
with peak luminosities between that of normal and superluminous SNe have also
been discovered \citep{Arcavi16}. Like normal SNe, SLSNe and luminous SNe
can also be classified into types I and II and most of SLSNe I are SNe Ic.

Previous research have shown that the LCs of almost all ordinary SNe can be
explained by the $^{56}$Ni model and the hydrogen recombination model while
the LCs of almost all SLSNe and luminous SNe cannot be explained by the $^{56}$Ni model.
Alternatively, the magnetar spin-down model \citep{Kasen10,Inserra13,Nicholl13,Nicholl14,Wang15,Wanglj16},
the ejecta-circumstellar medium (CSM) interaction (CSI) model
\citep{Chevalier82,Chevalier94,Chevalier11,Chatzopoulos12,Liu18}, and the fallback model \citep{Dexter13}
have been employed to account for the LCs of SLSNe, some luminous SNe as well as
a fraction of rapid rising luminous optical transients (see, \citealt{Moriya18,Wang19,Inserra19}
for reviews and references therein).

Nevertheless, all energy-source models mentioned above (the $^{56}$Ni model, the magnetar model,
the CSI model and the fallback model) cannot reproduce the LCs of SNe showing two or more peaks,
(e.g. iPTF13ehe, \citealt{Yan15}; iPTF15esb, \citealt{Yan17}; PTF12dam, and iPTF13dcc, \citealt{Vreeswijk17}).
To resolve this problem, various models have been developed.
\cite{Piro15} develop a semi-analytic post-shock cooling model that can be used
to account for the first peak of the LC of a double-peaked SN and the second peak of the
SN can be explained by the aforementioned models.
\cite{Wang16} propose that a double energy source model (magnetar plus CSI)
or a triple energy source model ($^{56}$Ni plus magnetar plus CSI) can be used to
explain the LCs of SLSNe exhibiting a rebrightening or double-peaked feature;
\cite{Vreeswijk17} suggest that a cooling plus magnetar model can explain the
LCs of PTF12dam and iPTF13dcc; \cite{Liu18} develop a multiple CSI model involving
the collisions between the ejecta and the CSM shells or winds locating at different
sites and use these models to fit the LCs of iPTF15esb and iPTF13dcc.

In this paper, we study two hydrogen-poor SNe, PS1-12cil and SN 2012aa.
PS1-12cil is a SLSN I discovered by the Panoramic Survey Telescope $\&$
Rapid Response System (Pan-STARRS1, PS1; \citealt{Kaiser10,Tonry12})
Medium Deep Survey (PS1 MDS) at a redshift ($z$) of 0.32 \citep{Lunnan18};
SN 2012aa is a luminous SN Ic that was discovered by the Lick Observatory Supernova
Search (LOSS, \citealt{Filippenko01}) at a redshift ($z$) of 0.083 \citep{Roy16}.
The multi-band LCs of these two SNe show unusual post-peak bumps \citep{Lunnan18,Roy16}
and make them double-peaked SNe which cannot be explained by any models that can reproduce
single-peak LCs.
We research the possible energy sources that can account for the double-peaked LCs
of PS1-12cil and SN 2012aa and constrain the physical parameters of the models.
In Section \ref{sec:fit}, we use the $^{56}$Ni plus CSI model,
the magnetar plus CSI model, and the multiple CSI model to fit the bolometric
LCs of PS1-12cil and SN 2012aa and derive the best-fitting parameters.
Our discussion and conclusions can be found in Sections \ref{sec:dis}
and \ref{sec:con}, respectively.

\section{Modeling the Bolometric LCs of PS1-12cil and SN 2012aa}
\label{sec:fit}

In their paper, \cite{Roy16} employ the $^{56}$Ni model and
the magnetar model to model the bolometric LC of SN 2012aa. It can be
found that while the first peak and the late-time LC of SN 2012aa
can be well fit by these two models, the second peak, i.e., the post-peak
bump cannot be explained by these models. More complicated models are
needed to resolve this problem. {\cite{Roy16} suggest that $^{56}$Ni
plus CSI model might be a reasonable model to account for the whole
double-peaked LC of SN 2012aa, but they don't model the LC using this model.}
Similarly, the double-peaked LC of PS1-12cil
must be modelled by more complicated models rather than simpler models like
the $^{56}$Ni model and the magnetar model. Therefore, we don't consider
these simple models throughout this paper.

In this section, we reproduce the bolometric LCs of PS1-12cil and SN 2012aa
by adopting the $^{56}$Ni plus CSI model, the magnetar plus CSI model,
and the multiple CSI model. Markov Chain Monte Carlo (MCMC) technique is
employed to obtain the best-fitting parameters and the 1$\sigma$ uncertainties.

\subsection{The $^{56}$Ni Plus CSI Model}

Firstly, we employ the $^{56}$Ni plus CSI model to fit the bolometric LCs of PS1-12cil
and SN 2012aa. We suppose that the $^{56}$Ni power the first peaks, while the
second peaks (bumps) can be explained by the CSI model. The $^{56}$Ni model in our
double-energy-source model including following parameters: the optical opacity $\kappa$,
the ejecta mass $M_{\rm{ej}}$, the mass of $^{56}$Ni $M_{\mathrm{Ni}}$, the gamma-ray
opacity of of $^{56}$Ni decay photons $\kappa_{\gamma,\mathrm{Ni}}$, the moment of
explosion $t_{\rm{expl}}$, the velocity of the ejecta $v_{\rm{SN}}$.
Based on their photospheric velocity from \citep{Lunnan18} and \citep{Roy16},
the velocities of the ejecta of PS1-12cil and SN 2012aa are
set to be $15,000\ \rm{km\ s^{-1}}$ and $11,400\ \rm{km\ s^{-1}}$,
respectively.

The density profile of the inner regions and the outer regions of the ejecta
can be described by $\rho_{\rm{ej}} \propto R_{\rm ej}^{-\delta}$ and
$\rho_{\rm{ej}} \propto R_{\rm ej}^{-n}$ (here, we take $\delta = 0$ and $n = 7$),
respectively; the density profile of the CSM can be described by
$\rho_{\rm CSM} \propto R_{\rm CSM}^{-s}$
($s = 0$ for material shells, $s = 2$ for stellar winds).
Forward shock and reverse shock generated by the interactions between the ejecta
and CSM propagate into CSM and ejecta, respectively, converting the kinetic energy
to radiation and driving the LCs of the bumps.

The parameters of the CSI model are the CSM mass $M_{\rm{CSM}}$,
the density of the innermost part of the CSM $\rho_{\rm{CSM,in}}$,
the efficiency of conversion from the kinetic energy
to radiation $\epsilon$, the dimensionless position parameter of break in
the ejecta from the inner region to the outer region $x_{\rm{0}}$, the
trigger moment of the interactions $t_{\rm{CSI}}$, and $M_{\rm{ej}}$.

\begin{figure}[tbph]
\begin{center}
\includegraphics[width=0.45\textwidth,angle=0]{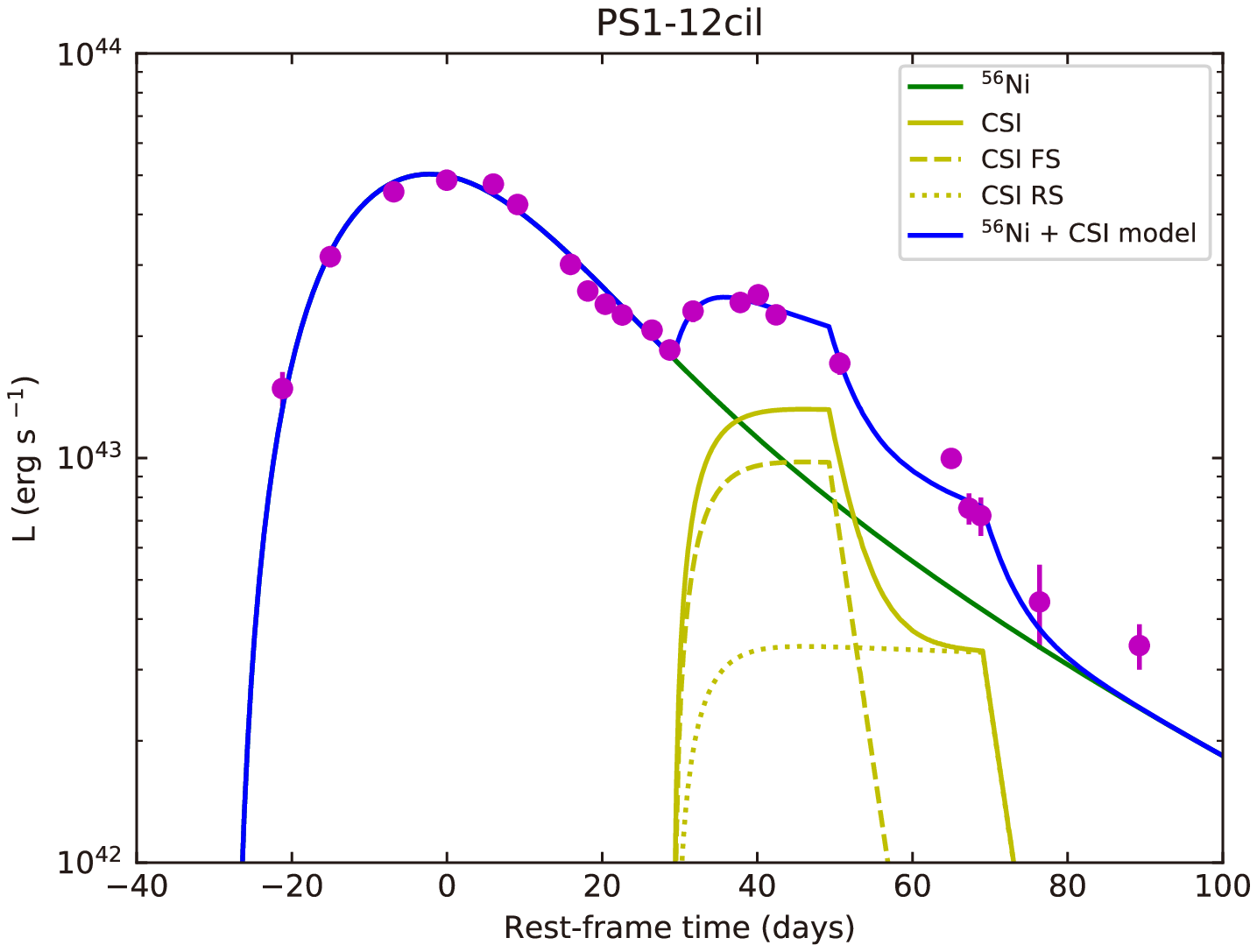}
\includegraphics[width=0.45\textwidth,angle=0]{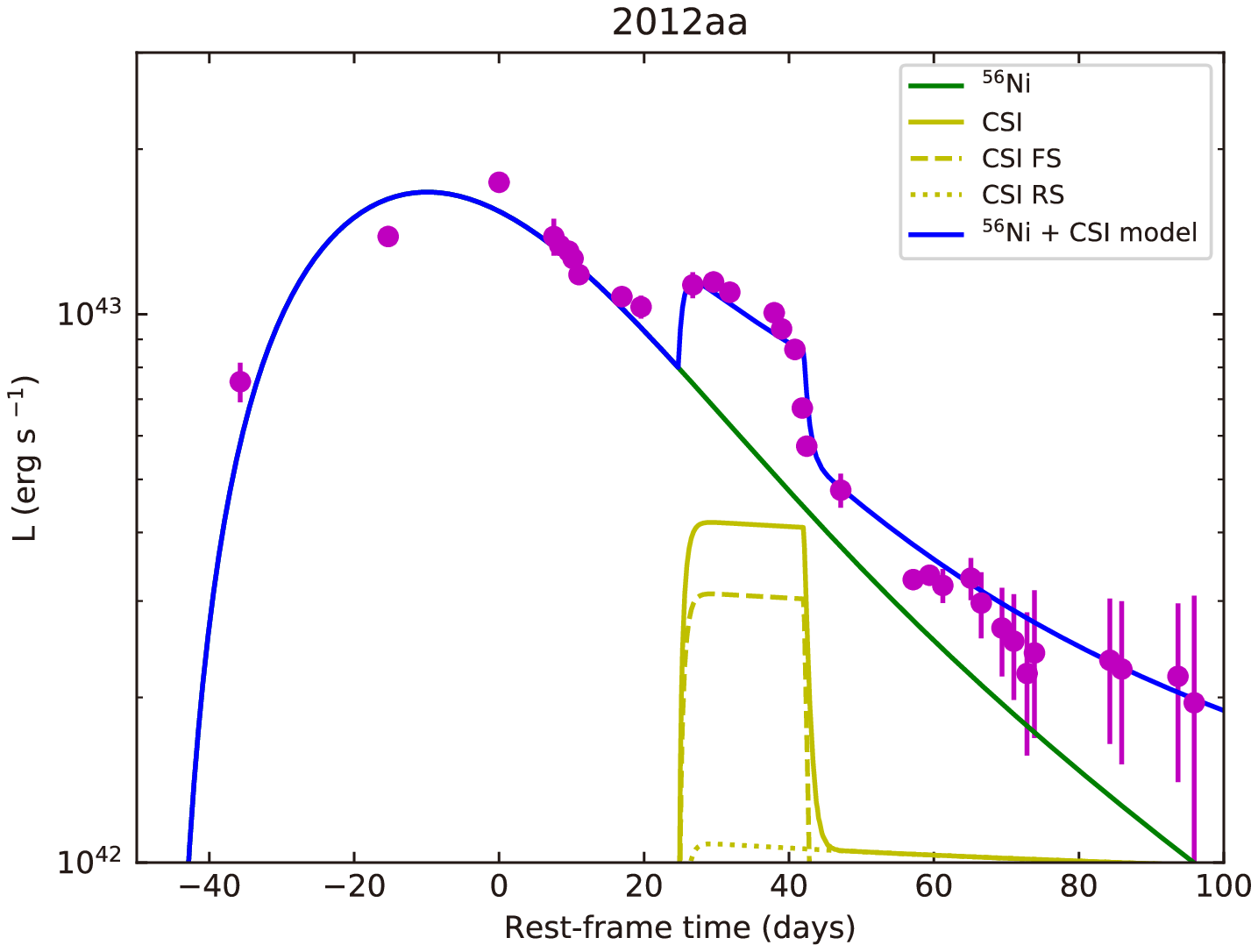}
\end{center}
\caption{The bolometric LCs of PS1-12cil (left panel) and SN 2012aa (right panel) reproduced by the $^{56}$Ni plus CSI model.
Data are taken from \cite{Lunnan18} and \cite{Roy16}, respectively.
The abscissa represents time since the explosion in the rest frame.}
\label{fig:Ni+CSI}
\end{figure}

The theoretical bolometric LCs reproduced by the $^{56}$Ni plus CSI model
are shown in Figure \ref{fig:Ni+CSI}. It can be found that the theoretical
LCs can match the observational data.
For PS1-12cil, the masses of the ejecta, $^{56}$Ni, and CSM are
$M_{\rm{ej}} = 4.95_{-0.73}^{+0.76} \rm\ M_{\odot}$,
$M_{\rm{Ni}} = 3.57_{-0.06}^{+0.07} \rm\ M_{\odot}$,
$M_{\rm{CSM}} = 4.66_{-0.70}^{+0.76} \rm\ M_{\odot}$, respectively;
for SN 2012aa, the masses of the ejecta, $^{56}$Ni, and CSM are
$M_{\rm{ej}} = 5.89_{-0.12}^{+0.12} \rm\ M_{\odot}$,
$M_{\rm{Ni}} = 1.63_{-0.01}^{+0.01} \rm\ M_{\odot}$,
$M_{\rm{CSM}} = 1.47_{-0.09}^{+0.15} \rm\ M_{\odot}$.
These three parameters and other best-fit parameters are listed
in Table \ref{tab:Ni+CSI}.

\subsection{The Magnetar Plus CSI Model}

Replacing the mass of $^{56}$Ni $M_{\mathrm{Ni}}$ and the gamma-ray
opacity of of $^{56}$Ni decay photons $\kappa_{\gamma,\mathrm{Ni}}$ by
the magnetic field strength $B_p$, the initial spin periods $P_0$, and
the gamma-ray opacity of magnetar spinning-down generated photons
$\kappa_{\gamma,\rm{mag}}$, the $^{56}$Ni plus CSI model can be
changed to the magnetar plus CSI model. Here, we use this
model to fit the LCs of For PS1-12cil and SN 2012aa.

\begin{figure}[tbph]
\begin{center}
\includegraphics[width=0.45\textwidth,angle=0]{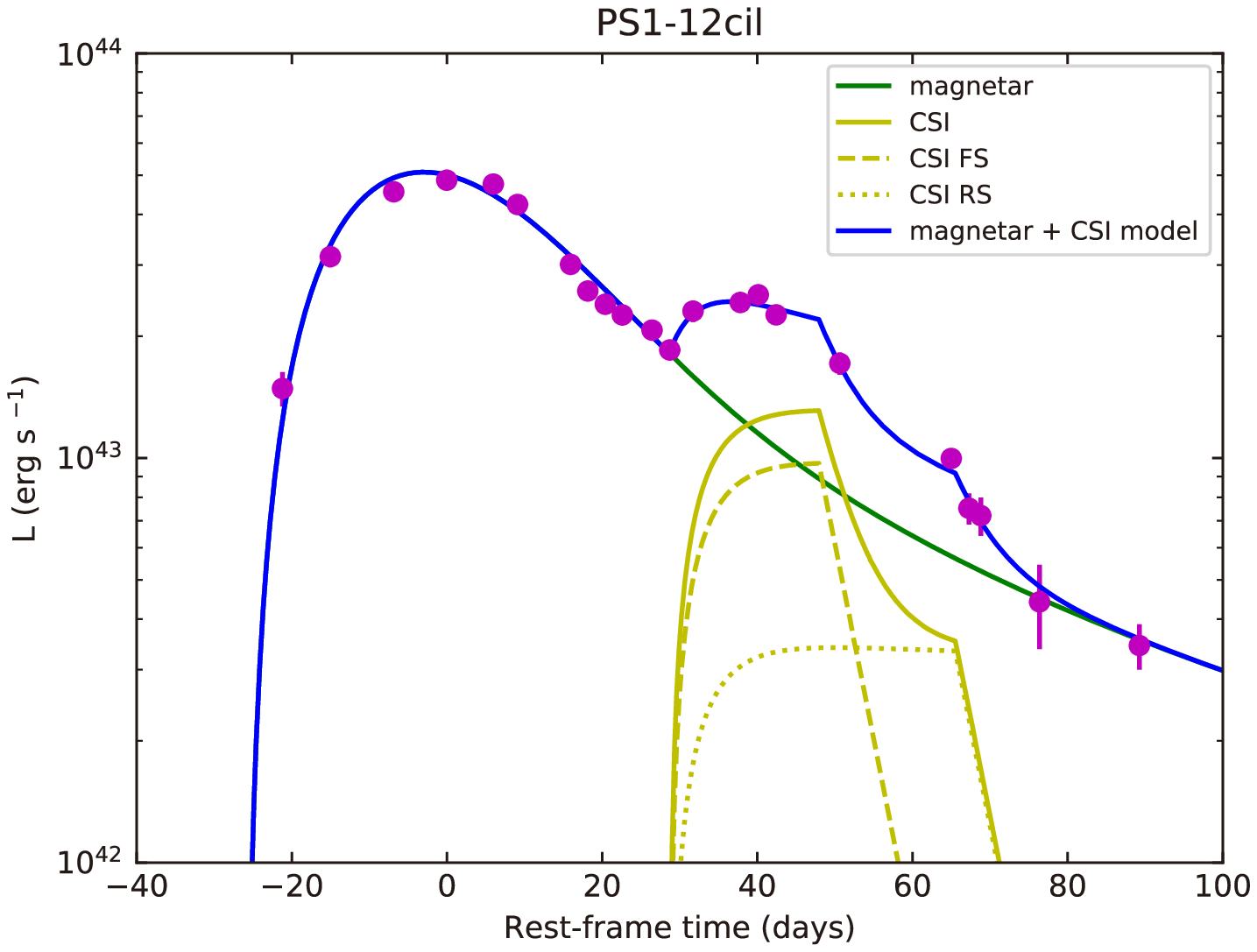}
\includegraphics[width=0.45\textwidth,angle=0]{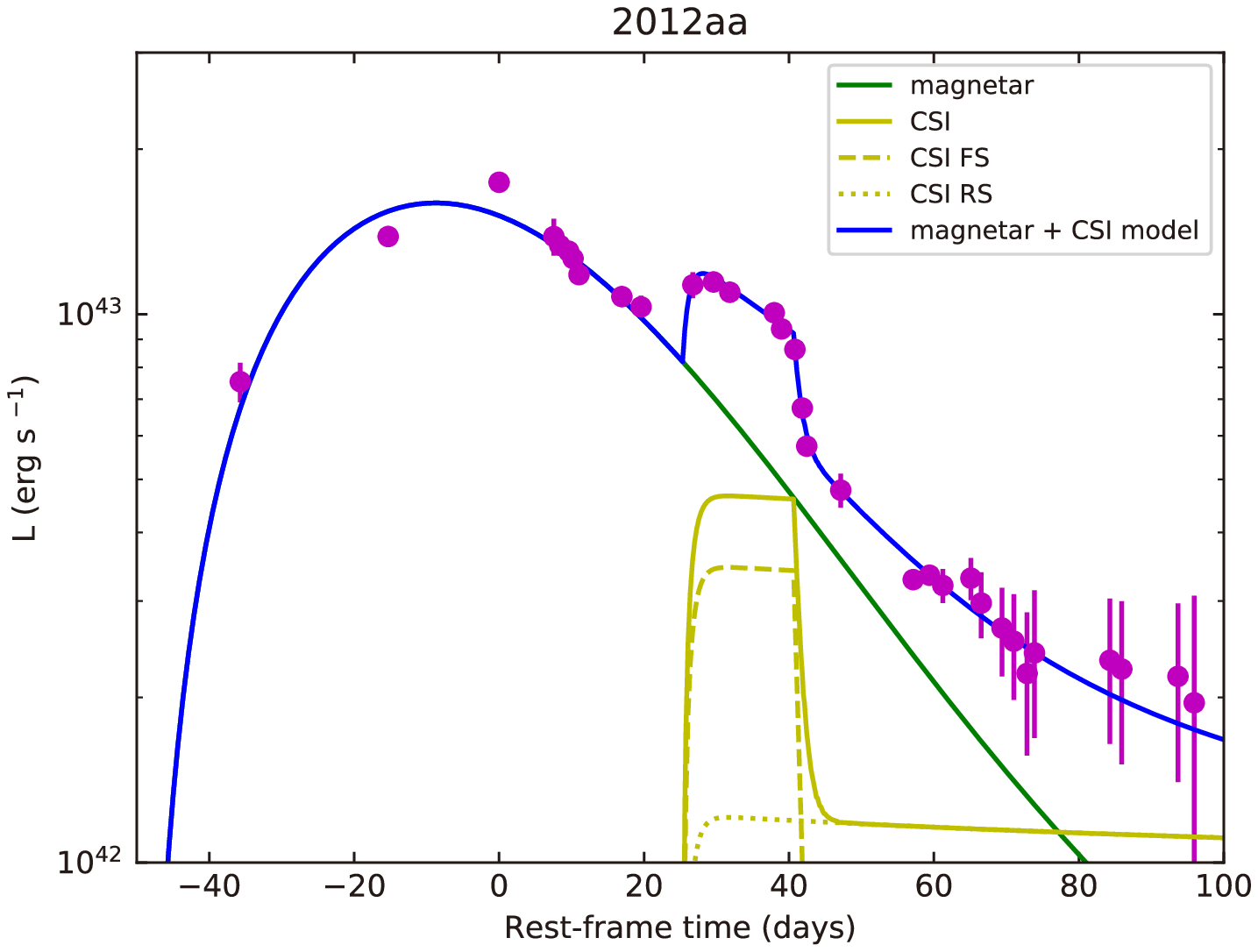}
\end{center}
\caption{The bolometric LCs of PS1-12cil (left panel) and SN 2012aa (right panel) reproduced by the magnetar plus CSI model.
Data are taken from \cite{Lunnan18} and \cite{Roy16}, respectively.
The abscissa represents time since the explosion in the rest frame.}
\label{fig:mag+CSI}
\end{figure}

The theoretical LCs yielded by the magnetar plus CSI model are plotted
in Figure \ref{fig:mag+CSI}. They also match the observational data.
For PS1-12cil, the best-fitting parameters are:
$M_{\rm{ej}} = 10.59_{-3.37}^{+3.25} \rm\ M_{\odot}$,
$P_0 = 6.52_{-0.10}^{+0.09} \rm\ ms$,
$B_p = 5.09_{-0.13}^{+0.13} \times 10^{14} \rm\ G$,
$M_{\rm{CSM}} = 9.64_{-3.04}^{+2.97} \rm\ M_{\odot}$;
for SN 2012aa, the best-fitting parameters are:
$M_{\rm{ej}} = 10.28_{-0.61}^{+0.72} \rm\ M_{\odot}$,
$P_0 = 8.99_{-0.14}^{+0.11} \rm\ ms$,
$B_p = 4.68_{-0.29}^{+0.40} \times 10^{14} \rm\ G$,
$M_{\rm{CSM}} = 1.93_{-0.12}^{+0.15} \rm\ M_{\odot}$.
These four parameters and other best-fit parameters are listed
in Table \ref{tab:mag+CSI}.

\subsection{The Multiple CSI Model}

The multiple CSI model developed by \cite{Liu18} is also a promising model that can
account for two or more LC peaks of SNe. The free parameters of multiple CSI model
are listed in \ref{tab:multiCSI}.
Based on the shapes of bolometric LCs of PS1-12cil and SN 2012aa,
we assume that there are two interactions between ejecta and CSM.
In other words, the multiple CSI model adopted here is a double CSI model.
It is reasonable to suppose that the outer CSM are shells ($s = 0$)
while the inner CSM can either be the stellar wind ($s = 2$) or
the shell ($s = 0$). Both these two combinations are considered here.

\begin{figure}[tbph]
\begin{center}
\includegraphics[width=0.45\textwidth,angle=0]{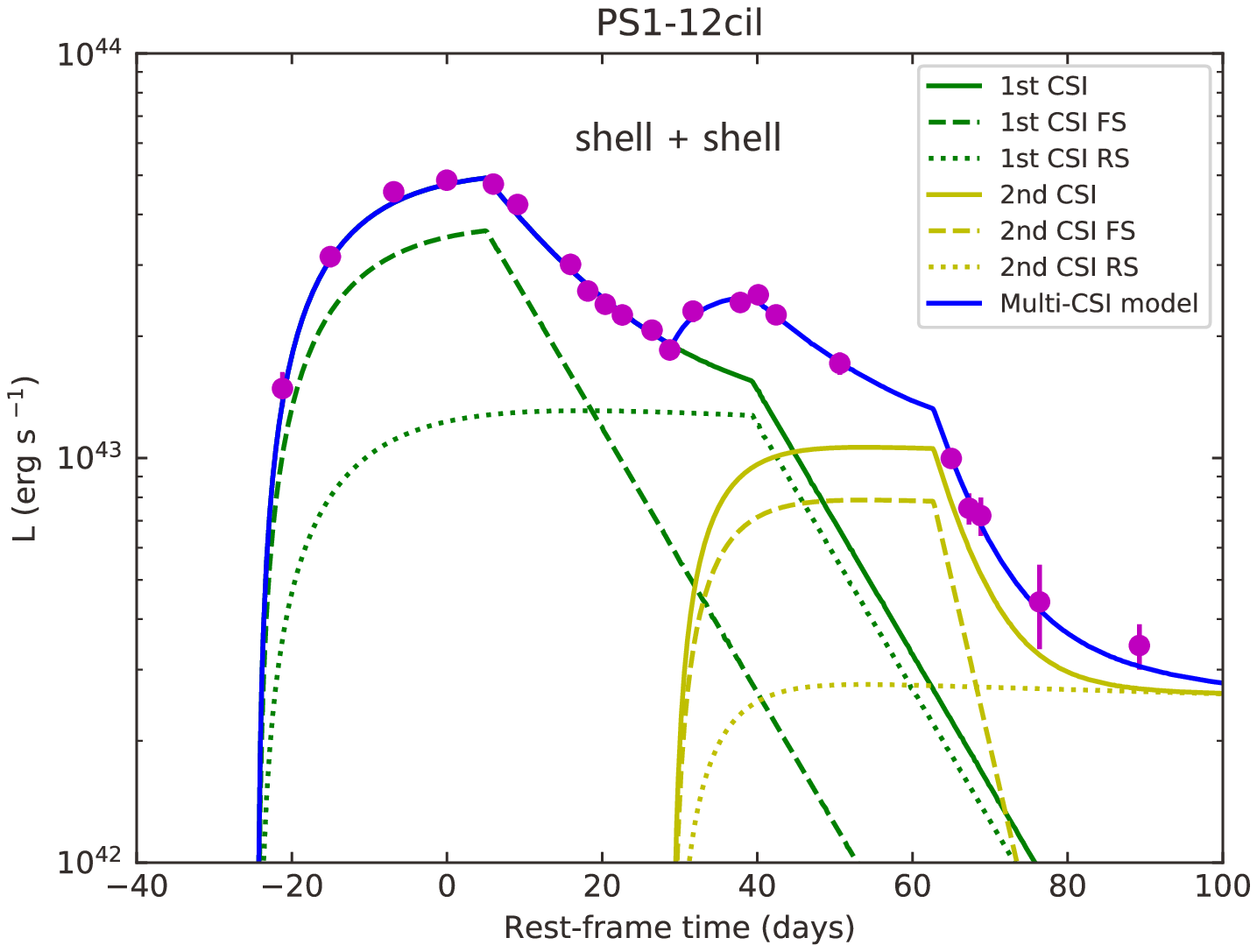}
\includegraphics[width=0.45\textwidth,angle=0]{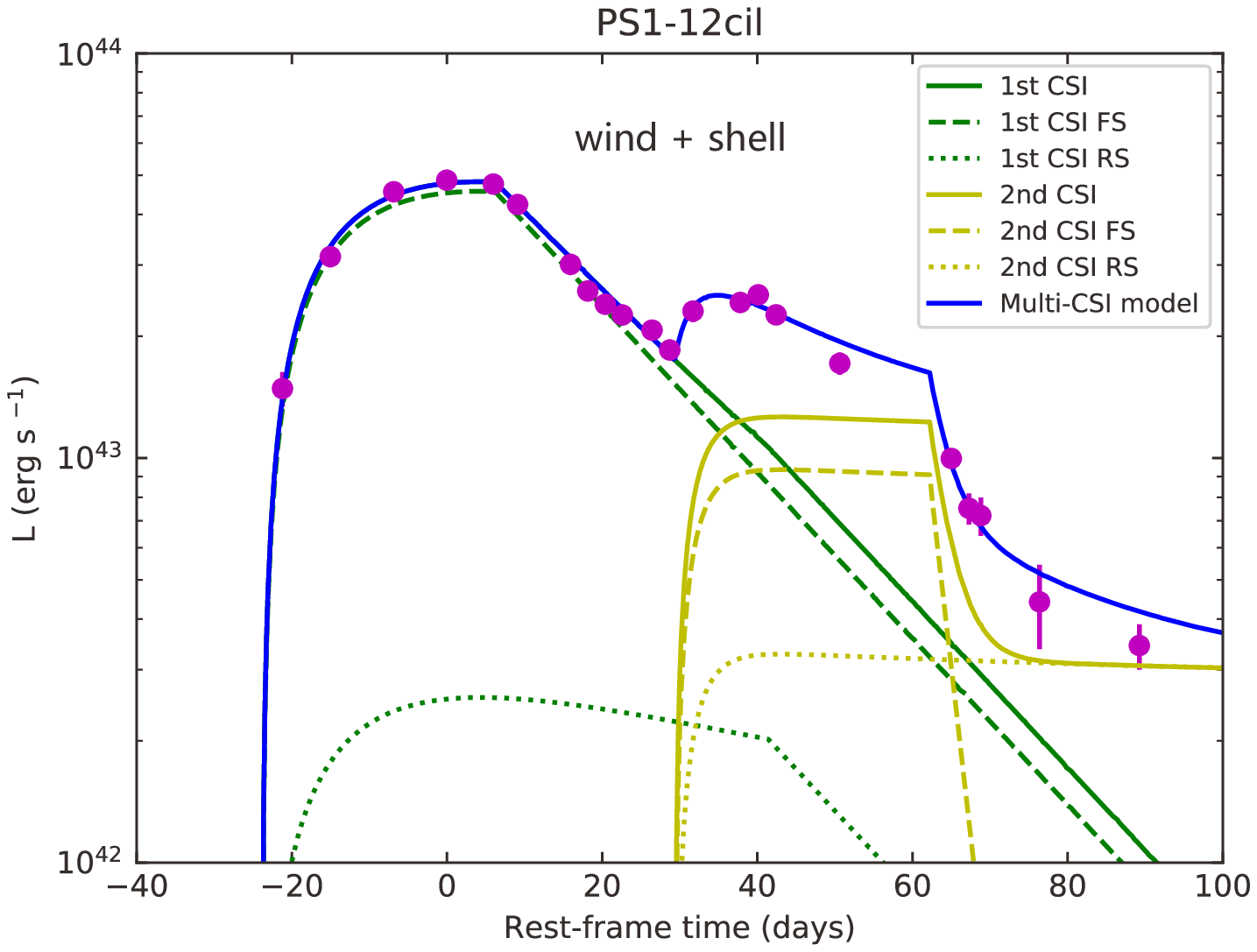}
\includegraphics[width=0.45\textwidth,angle=0]{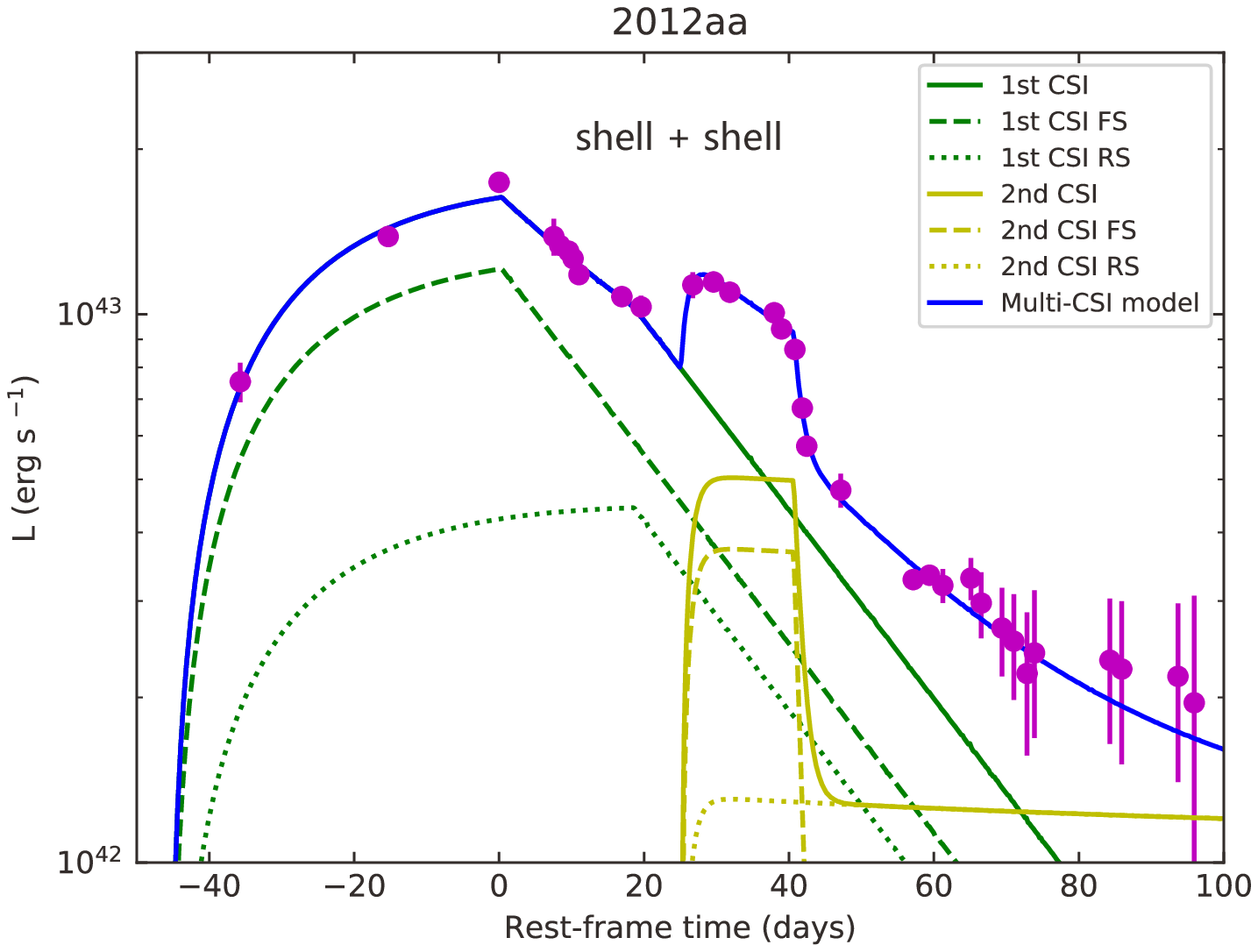}
\includegraphics[width=0.45\textwidth,angle=0]{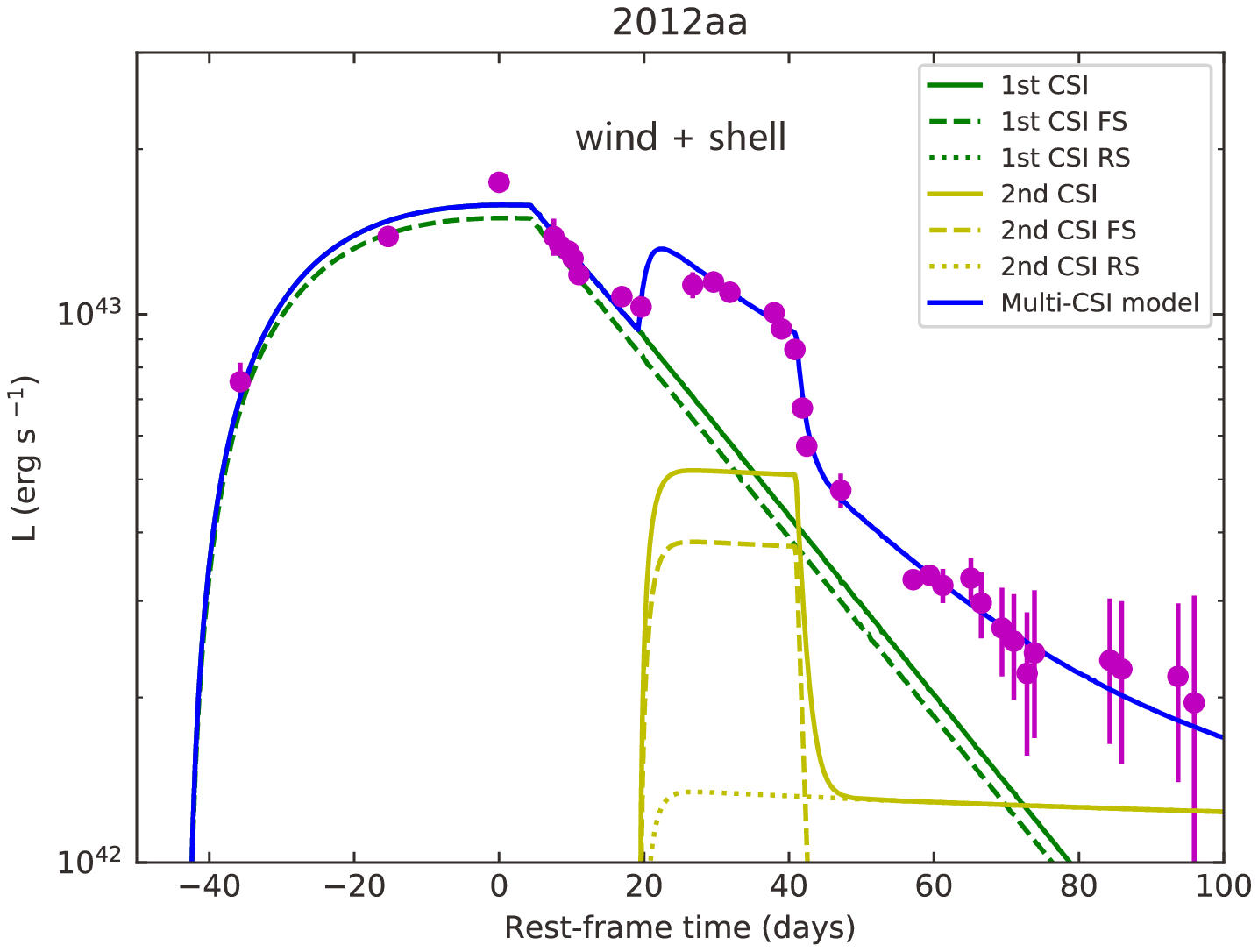}
\end{center}
\caption{The bolometric LCs of PS1-12cil (top panels) and SN 2012aa (bottom panels)
reproduced by the double CSI model. The left panels represent the double shell ($s=0$) case,
while the right panels represent the wind ($s=2$) plus shell ($s=0$) case.
Data are taken from \cite{Lunnan18} and \cite{Roy16}, respectively.
The abscissa represents time since the explosion in the rest frame.}
\label{fig:multiCSI}
\end{figure}

The LCs produced by the double CSI model are shown
in Figure \ref{fig:multiCSI} and the best-fitting parameters are listed
in Table \ref{tab:multiCSI}.
For PS1-12cil, $M_{\rm{ej},1} = 25.64_{-3.12}^{+2.51} \rm\ M_{\odot}$,
$M_{\rm{CSM},1} = 17.19_{-1.88}^{+1.59} \rm\ M_{\odot}$,
$M_{\rm{CSM},2} = 4.07_{-0.51}^{+0.65} \rm\ M_{\odot}$ if
the inner CSM is a wind ($s = 2$); otherwise ($s = 0$),
$M_{\rm{ej},1} = 13.18_{-2.07}^{+2.56} \rm\ M_{\odot}$,
$M_{\rm{CSM},1} = 7.13_{-1.15}^{+1.39} \rm\ M_{\odot}$,
$M_{\rm{CSM},2} = 6.26_{-0.97}^{+0.93} \rm\ M_{\odot}$.
For SN 2012aa, $M_{\rm{ej},1} = 19.46_{-1.25}^{+1.70} \rm\ M_{\odot}$,
$M_{\rm{CSM},1} = 19.41_{-0.65}^{+0.37} \rm\ M_{\odot}$,
$M_{\rm{CSM},2} = 2.36_{-0.17}^{+0.15} \rm\ M_{\odot}$ if
the the inner CSM is a wind ($s = 2$); otherwise ($s = 0$),
$M_{\rm{ej},1} = 15.60_{-0.97}^{+0.93} \rm\ M_{\odot}$,
$M_{\rm{CSM},1} = 17.77_{-0.96}^{+0.95} \rm\ M_{\odot}$,
$M_{\rm{CSM},2} = 2.16_{-0.19}^{+0.20} \rm\ M_{\odot}$.

\section{Discussion}
\label{sec:dis}

\subsection{Which is the Best Model?}

\subsubsection{PS1-12cil}

For PS1-12cil, the $^{56}$Ni plus CSI model is disfavored since the
ratio of the mass of $^{56}$Ni ($3.57_{-0.06}^{+0.07} \rm\ M_{\odot}$)
to the mass of the eject ($4.95_{-0.73}^{+0.76} \rm\ M_{\odot}$)
$M_{\rm{Ni}}/M_{\rm{ej}}$ is $\sim 0.72$ while the upper limit of this
ratio is $\sim 0.2$ \citep{Umeda08} and \cite{Inserra13} suggest that
0.5 can be regarded as the upper limit of this ratio.

{The theoretical LC yielded by the magnetar plus CSI model can
match the data well and the value of $\chi^2/\rm{dof}$ of this model
is $54.39/11=4.94$.}

{For the two shell CSI model, {the masses of} the inner shell
($M_{\rm{CSM},1}$) and the outer shell ($M_{\rm{CSM},2}$) are
$\sim 7.13 \rm\ M_{\odot}$) and $\sim 6.26 \rm\ M_{\odot}$, respectively;
the value of $\chi^2/\rm{dof}$ of this model is $21.08/10=2.11$.}

For the case of the ``inner wind plus outer shell" combination,
{the value of $\chi^2/\rm{dof}$ of this model
is $31.48/10=3.15$}. Assuming that the velocity of the wind {of the
progenitor of an SN Ibc} is
($v_{\rm w}$) is $100-1000\rm\ km\ s^{-1}$
(see, e.g., Table 1 of \citealt{Smith2014ARA&A..52..487S}),
the mass loss rate $\dot M = 4 \pi v_{\rm w} q$
($q = \rho_{\rm CSM,in,1} R_{\rm CSM,in,1}^{2}$)
of the stellar wind is $\dot M \sim 10.1-101\ \rm M_\odot\ yr^{-1}$,
significantly larger than the typical values for most of SN progenitors.
\footnote{For instance, the values of $\dot M$ of SN~1994W, SN~1995G,
and iPTF13z are $\sim 0.2\ \rm M_\odot\ yr^{-1}$ \citep{Chugai04},
$\sim 0.1\ \rm M_\odot\ yr^{-1}$ \citep{Chugai03}, and
$\sim 0.1-2\ \rm M_\odot\ yr^{-1}$ \citep{Nyholm17}, respectively.
{\citet{Smith2013MNRAS.429.2366S} suggest that the CSI between the ejecta
($M_{\rm ej} \sim 10\rm M_\odot$ and a srong wind ($\dot M \sim 0.33\ \rm M_\odot\ yr^{-1}$)
blowing for 30 years could power the light curve of the 19th century eruption of Eta Carinae.
\citet{Ofek2014ApJ...781...42O} propose that the interaction between the ejecta and a wind
with mass loss rate $\dot M \approx 0.8(v_{\rm CSM}/300~\rm km~s^{-1})\ \rm M_\odot\ yr^{-1}$
can account for the light curve of SN~2010jl
\citep{Smith2011ApJ...732...63S,Stoll2011ApJ...730...34S,Smith2012AJ....143...17S,
Zhang2012AJ....144..131Z,Ofek2014ApJ...781...42O,Fransson2014ApJ...797..118F}
and the accumulated CSM over $\sim 10$ or 16 years is $\sim 10 \rm M_\odot$}.}

{To produce such extreme wind mass loss, so-called ``super-winds"
are required. Recently, \cite{Moriya19} suggest that  the multi-peaked
IIP SN iPTF14hls is a optical transient powered by mass loss history.
In this scenario, the maximum mass loss rate must be larger than
10 $\rm M_{\odot}$ yr $^{-1}$. As pointed out by \cite{Moriya19}, however,
how to produce so large mass loss is unclear. Moreover, the velocity
of the putative wind of PS1-12cil might be significantly larger than
100 $\rm km~s^{-1}$ and can be up to 1000 $\rm km~s^{-1}$ since
PS1-12cil is a hydrogen-poor SLSN whose progenitor might be a Wolf-Rayet-like
star which would blow high-velocity wind; in this case, the mass loss rate
would be significantly larger than 10 $\rm M_{\odot}$ yr $^{-1}$. Therefore,
the scenario involving a wind with mass loss rate $\gtrsim$ 10 $\rm M_{\odot}$ yr $^{-1}$
need some unknown extraordinary mechanism.}

Comparing the $\chi^2/\rm{dof}$ of these models,
{the double CSI model involving two shells ($s = 0$)} is the best one
for PS1-12cil, {while the magnetar plus CSI model is also a promising model
that can account for the LC of PS1-12cil}.

{The spectral features would provide more information. 
The PS1-12cil's spectrum obtained 1 day after the peak don't show emission line 
due to the the CSI, indicating either that the early LC of PS1-12cil might be
powered by the magnetar rather than CSI or that the CSM is asymmetric and the
emission lines were swallowed by the ejecta. There are no late-time spectra and 
the power sources powering the late-time LC cannot be constrained by spectra.
If we neglect the asymmetry effect, the magnetar plus CSI model is the best one.}

\subsubsection{SN 2012aa}

For SN 2012aa, {the value of $\chi^2/\rm{dof}$ of the
$^{56}$Ni plus CSI model is $92.90/22=4.22$.} The model need
$1.63_{-0.01}^{+0.01} \rm\ M_{\odot}$ of $^{56}$Ni;
the ejecta mass derived is $5.89_{-0.12}^{+0.12} \rm\ M_{\odot}$,
indicating that the value of $M_{\rm{Ni}}/M_{\rm{ej}}$ is
$\sim 0.28$. This value is larger
than 0.2 but smaller than 0.5, so the $^{56}$Ni plus CSI model
cannot be completely excluded.

{Based on the parameters derived, the value of the kinetic
energy of SN 2012aa can be calculated, $\sim 4.6\times 10^{51}$ erg,
which is lower than those of some hypernovae ($\gtrsim 10^{52}$ erg,
e.g., SNe 1998bw and 2003dh),
suggesting that the mass of $^{56}$Ni must be lower than those of these hypernovae
for which the inferred $^{56}$Ni masses are $\lesssim 0.5 \rm\ M_{\odot}$.
\footnote{Larger kinetic energy would produce more $^{56}$Ni, see
Fig 8 of \citet{Mazzali2013MNRAS.432.2463M},
\citet{2015ApJ...807..147W} and references therein.}
However, the $^{56}$Ni mass derived by $^{56}$Ni plus CSI model is 
$1.63_{-0.01}^{+0.01} \rm\ M_{\odot}$, significantly higher than
those of SNe 1998bw and 2003dh. So this model is disfavored.}

The values of the parameters of the magnetar plus CSI model
are reasonable, and the value of $\chi^2/\rm{dof}$ of this model is $36.77/21=1.75$.
For the two shell CSI model, {the masses of} the inner shell
($M_{\rm{CSM},1}$) and the outer shell ($M_{\rm{CSM},2}$) are
$\sim 18 \rm\ M_{\odot}$) and $\sim 2 \rm\ M_{\odot}$, respectively;
{the value of $\chi^2/\rm{dof}$ of this model is $18.61/20=0.93$.}
For wind plus shell CSI model,
the derived value of $\dot M$ is $\sim 10.4-104\ \rm M_\odot\ yr^{-1}$,
{and the value of $\chi^2/\rm{dof}$ of this model is $44.23/20=2.21$.}

So {the two shell CSI model} is the best one for SN 2012aa and
{the magnetar plus CSI model is also a reasonable model in explaining
the SN 2012aa's LC}.

{The early spectra of SN 2012aa resemble those of SNe Ic-BL,
\citet{Roy16} point out that the broad emission feature (Gaussian FWHM
$\approx 14,000 \rm~km~s^{-1}$) near H$\alpha$ remains almost constant
throughout its evolution. This feature is prominent in the spectrum 47 d
after the LC peak of SN 2012aa. \citet{Roy16} suggest that this feature could be due to [O I]
$\lambda\lambda$6300, 6364 or blueshifted (1500-2500 $\rm km s^{-1}$) H$\alpha$
emission.}

For the latter explanation, \citet{Roy16} propose that the feature is
consistent with the CSM-interaction scenario since it becomes prominent by +47 d when the
LC peaks again. If the early-time LC was also powered by CSI, the possible
emission lines associated with CSI might be stronger than that in the late-time
spectra since the density and mass of the inner shell are larger than those of
outer shell, inconsistent with the observations which show that
the feature was marginally present in the spectrum by +29 d and become prominent
in the spectrum +47 d. Therefore, the magnetar plus CSI model is favored.

Nevertheless, if the broad emission feature is [O I]
$\lambda\lambda$6300, 6364 rather than the CSI-induced emission lines, we
can expect that the CSI between the ejecta and the asymmetric CSM could produce the CSI
without emission lines. In this case, both the two shell CSI model and the magnetar
plus CSI model are plausible.

\subsection{The the Mass Loss Histories of PS1-12cil and SN 2012aa}

The most promising models of PS1-12cil and SN 2012aa is {the two shell CSI model}.
However, the magnetar plus CSI model is also a possible model for these two SNe.
It is interesting to infer the mass-loss histories of these two SNe {based on
these two models}.

Provided that ${\Delta}t'$ is the interval between the time when the shells were
expelled and the time when the shells were collided,
${\Delta}t=t_\mathrm{CSI}-t_\mathrm{expl}$ is the
interval between the time when the SNe exploded
and the time when the shells were collided, we have
$v_{\rm SN}{\Delta}t=v_{\rm shell}({\Delta}t'+{\Delta}t)$,
i.e., ${\Delta}t'=(v_{\rm SN}/v_{\rm shell}-1){\Delta}t$,
here, $v_{\rm shell}$ is the velocity of the shells.
Supposing that the value of the velocity of the shells
is $100-1000\rm\ km\ s^{-1}$, we can infer the values of $t'$.


{Under the hypothesis that the LCs of PS1-12cil and SN 2012aa
were powered by two shell CSI, the values of of $t'$ can be
derived below. For PS1-12cil, $v_{\rm SN}=15,000\ \rm{km\ s^{-1}}$,
${\Delta}t_{\rm inner} \sim 8.70$ days, ${\Delta}t_{\rm inner}' \sim 121.74-1295.70$ days
($\sim 0.3-3.6$ years); ${\Delta}t_{\rm outer} \sim 62.14$ days,
${\Delta}t_{\rm outer}' \sim 869.90-9258.26$ days ($\sim 2.4-25.4$ years).
For SN 2012aa, $v_{\rm SN}=11,400\ \rm{km\ s^{-1}}$,
${\Delta}t_{\rm inner} \sim 8.27$ days, ${\Delta}t_{\rm inner}' \sim 86.05-935.01$ days
($\sim 0.2-2.6$ years); ${\Delta}t_{\rm outer} \sim 79.08$ days,
${\Delta}t_{\rm outer}' \sim 822.48-8936.54$ days ($\sim 2.3-24.5$ years).
We find that the values of ${\Delta}t'$ of the inner shells of these two SNe
are approximately equal to each other; similarly, the values of ${\Delta}t'$
of the outer shells of these two SNe are also approximately equal to each other.}


{Similarly, we can infer the values of of $t'$ based
on the assumption that these two SNe were powered by magnetars
and CSI. For PS1-12cil, $v_{\rm SN}=15,000\ \rm{km\ s^{-1}}$,
${\Delta}t \sim 55$ days, then ${\Delta}t' \sim 770-8200$ days,
$\sim 2-22$ years. For SN 2012aa, $v_{\rm SN}=11,400\ \rm{km\ s^{-1}}$,
${\Delta}t \sim 75$ days, then ${\Delta}t' \sim 780-8500$ days,
$\sim 2-23$ years. The values of ${\Delta}t'$ of these two SNe
are equal to each other.}

\subsection{Comparing the Properties of these two SNe and other double-peaked SNe associated with CSI}

{It is interesting to compare the observation properties and the
derived physical parameters of these two SNe. From the observational aspect,
both two SNe have bright main-peaks, and the delay time between the
main-peaks and the bumps (second peaks) are about 40 days, indicating that
the model parameters might be rather similar if they share the same origin.}

{If we suppose that the first peaks of these two SNe were powered by the
magnetars, the initial spin periods ($P_0$) and the magnetic field strength
($B_p$) of the putative magnetars are $\sim7-10$ ms and $\sim5\times10^{14}$ G,
respectively. These values are consistent with that of the magnetars supposed to
power the LCs of type I SLSNe \citep{Liu2017ApJ...842...26L,Yu2017ApJ...840...12Y,
Nicholl2017ApJ...850...55N}. Moreover, the CSM shells should be expelled $\sim2-22$
years prior to the SN explosions if the post-peak bumps were powered by the CSI.}

{Provided that the whole LCs of PS1-12cil and SN 2012aa were powered by two
shell CSI, we find that the mass loss histories of inner shells ($\sim0.3-3$ years
prior to the SN explosions) and the outer shells ($\sim2-25$ years prior
to the SN explosions) of these two SNe are also approximately equal to each other.}

{The resemblances between these two SNe suggest that the properties of their
progenitors are similar to each other: they yield magnetars with nearly same properties
and experience the same pre-SN eruptions for in the magnetar plus CSI scenario, or
experience the two pre-SN eruptions sharing the properties consistent with each other.}

{Furthermore, we can compare these two SNe with other double-peaked SNe which
have $\sim 40$ days delay between two peaks. \citet{Smith2014} suggest that both
SN~2009ip (2012a/b, \citealt{Mauerhan2013MNRAS.430.1801M,
Pastorello2013ApJ...767....1P,Margutti2014ApJ...780...21M,Smith2010AJ....139.1451S})
and SN 2010mc \citep{Ofek2013Natur.494...65O,Smith2014MNRAS.438.1191S} are double-peaked
SNe whose second peaks are powered by CSI. \citet{Smith2014} suggest that the first peaks with
absolute magnitudes $\sim -15$ mag were powered by the $^{56}$Ni decay
while the second peaks $\sim 3$ mag brighter than the first
peaks were powered by CSI. The delay between the two first and second peaks
of the two SNe studied are also $\sim$ 40 days,
nearly identical to those of PS1-12cil and SN 2012aa.
Supposing that the first peaks of PS1-12cil and SN 2012aa were powered by
magnetars and the the first peaks of SN~2009ip and SN 2010mc while the
second peaks of the four SNe were powered by CSI, the resemblances
between the observations of these four SNe indicate that they might have similar
pre-SN mass loss histories.}

{While the delay between the first and second peaks of these four SNe
are $\sim 40$ days, there are some discrepancy between their observation features.
First, the first peaks of SN~2009ip and SN 2010mc are $\sim 3$ magnitude
dimmer than the second peaks; in contrast, the first peaks of PS1-12cil and SN 2012aa
are brighter than the second peaks. This discrepancy might be due to the different
energy sources: the first peaks of SN~2009ip and SN 2010mc have been believed to be
powered by $^{56}$Ni decay and the first peaks of PS1-12cil and SN 2012aa might be
powered by CSI or magnetars. Moreover, the r/R-band absolute magnitudes of
the first peaks of PS1-12cil and SN 2012aa are respectively $\sim -21$ mag
(derived by combining the Figure 2 of \citealt{Lunnan18} and PS1-12cil's redshift,
0.32) and $\sim -20$ mag \citep{Roy16}, at least $2-3$ magnitude brighter than
the first peaks of SN~2009ip and SN 2010mc.}

{Both SN~2012aa and SN~2009ip locate at the outskirts of their host galaxies
(the values for PS1-12cil and SN 2010mc have not been confirmed):
the distance between SN~2012aa and the galaxy center is $\sim$3 kpc, while
the distance between SN~2009ip and the galaxy center is $\sim 5$ kpc which is larger than
that of the former. The properties of the host galaxies and their location
are also rather different: the host of SN 2012aa is a normal star-forming galaxy with solar
metallicity, while the host of SN 2010mc is a faint low-metallicity dwarf galaxy.}

\subsection{Can the Cooling Emission Plus $^{56}$Ni/Magnetar Explain the LCs of PS1-12cil and SN 2012aa}
The post-shock cooling emission plus $^{56}$Ni/magnetar are also
plausible models that can account for some double-peaked SNe.
However, the cooling emission would power monotonically decreasing
bolometric LCs (see, e.g., \citealt{Piro15}) that don't have rising
parts. \footnote{The multi-band optical-IR LCs powered by
cooling emission have rising parts while the UV LCs are
monotonically decreasing ones.} In contrast, the LCs around the
first peaks of bolometric LCs of PS1-12cil and SN 2012aa shown clear rising phases.
Therefore, the models involving cooling emission are disfavored.

A shock breakout plus cooling emission may reproduce rising and
descending peak, but its rising phase is too short to compatible with
the bolometric LCs of PS1-12cil and SN 2012aa.

\section{Conclusions}
\label{sec:con}

PS1-12cil is a Type I SLSN, while SN 2012aa is a luminous Type Ic SN.
The bolometric LCs of these two SNe are double-peaked and cannot be
explained by any models that can reproduce single-peaked LCs.
In this paper, We use three models, the $^{56}$Ni plus CSI model,
the magnetar plus CSI model, and the double CSI model, to account for
their bolometric LCs.

According to the best-fitting parameters and the physical constraints,
we find that {the two shell CSI model} is the most
promising one that can reproduce the bolometric LCs of these two SNe.
For PS1-12cil, the model parameters are $M_{\rm{ej}} = 13.18_{-2.07}^{+2.56} \rm\ M_{\odot}$,
$M_{\rm{CSM,1}} = 7.13_{-1.15}^{+1.39} \rm\ M_{\odot}$,
and $M_{\rm{CSM,2}} = 6.26_{-0.97}^{+0.93} \rm\ M_{\odot}$
For SN 2012aa, the model parameters are $M_{\rm{ej}} = 15.60_{-0.97}^{+0.93} \rm\ M_{\odot}$,
$M_{\rm{CSM,1}} = 17.77_{-0.96}^{+0.95} \rm\ M_{\odot}$,
and $M_{\rm{CSM,2}} = 2.16_{-0.19}^{+0.20} \rm\ M_{\odot}$.

The magnetar plus CSI model {is also a good model to
fit the LCs of these two SNe.}
For PS1-12cil, the model parameters are
$M_{\rm{ej}} = 10.59_{-3.37}^{+3.25} \rm\ M_{\odot}$,
$P_0 = 6.52_{-0.10}^{+0.09} \rm\ ms$, $B_p = 5.09_{-0.13}^{+0.13} \times 10^{14} \rm\ G$, and
$M_{\rm{CSM}} = 9.64_{-3.04}^{+2.97} \rm\ M_{\odot}$.
For SN 2012aa, the model parameters are $M_{\rm{ej}} = 10.28_{-0.61}^{+0.72} \rm\ M_{\odot}$,
$P_0 = 8.99_{-0.14}^{+0.11} \rm\ ms$, $B_p = 4.68_{-0.29}^{+0.40} \times 10^{14} \rm\ G$, and
$M_{\rm{CSM}} = 1.93_{-0.12}^{+0.15} \rm\ M_{\odot}$.

Based on the explosion time $t_\mathrm{expl}$ and the collision time
$t_\mathrm{CSI}$ between the ejecta and the CSM, we can infer the mass
loss histories of the progenitors.
{Provided that they were powered by double ejecta-shell CSI,
the inner and outer shells were expelled $\sim 0.2-3.6$ and
$\sim 2-25$ years before the explosions of the SNe, respectively;}
{the shells were expelled} $\sim 2-20$ years before the explosions
of the SNe {if they were powered by magnetars plus CSI}.

The precursor eruption events have been observed (see, e.g., \citealt{Ofek14,Arcavi17})
in the images obtained several years or several decades before the detections
of the corresponding SNe II/IIn. Furthermore, \citet{Yan15} estimated that
at least 15$\%$ of SLSNe-I might expel precursor eruptions before the 
SN explosions; the interactions between SN ejecta and the CSM would
power a late-time LC rebrightening and the SNe would be confirmed as the double- or
multiple-peaked SNe if the late-time follow-up can be conducted successfully.

Researching the properties of the LCs of some double- or multi-peaked
SNe would provide useful information about the precursor eruptions.
It can be expected that future optical sky-survey telescopes would discover
more double- or multi-peaked SNe and the theoretical studies would
yield more valuable conclusions.

\acknowledgments
This work is supported by National Natural Science Foundation of China
(Grant Nos. 11963001 and 11533003), Guangxi Science Foundation (Grant Nos. 
2016GXNSFCB380005 and 2018GXNSFGA281007), and Bagui Young
Scholars Program (LHJ). Z.G.D. is supported by the National Key Research
and Development Program of China (grant No. 2017YFA0402600) and the National
Natural Science Foundation of China (grant Nos. 11573014 and 11833003). XGW
was supported by National Natural Science Foundation of China (grant No. 11673006),
the Guangxi Science Foundation (grant No. 2016GXNSFFA380006), the One-Hundred-
Talents Program of Guangxi colleges, and High level innovation team and outstanding
scholar program in Guangxi colleges. LDL is supported by  the National Postdoctoral
Program for Innovative Talents (grant No. BX20190044), China Postdoctoral Science
Foundation (grant No. 2019M660515) and ``LiYun'' postdoctoral fellow of Beijing
Normal University.


\bibliography{ms}

\clearpage

\begin{table*}[tbp]
\tabcolsep=2pt
\caption{Parameters of the $^{56}$Ni plus CSI model. The uncertainties are 1$\sigma$.}
\label{tab:Ni+CSI}
\begin{center}
{\scriptsize
\begin{tabular}{cccccccccccccccc}
\hline\hline																		
	&	$\kappa$	&	$M_{\mathrm{ej}}$	&	$M_{\mathrm{Ni}}$	&	$M_{\mathrm{CSM}}$	&	$\rho_{\mathrm{CSM,in}}$	&	$\epsilon$	&	$x_{\mathrm{0}}$	 &	$\kappa_{\gamma,\mathrm{Ni}}$	&	$t_\mathrm{expl}$	&	$t_\mathrm{CSI}$	&	$\chi^2/\mathrm{dof}$\\
	&	(cm$^2$ g$^{-1}$)	&	(M$_{\odot}$)	&	(M$_{\odot}$)	&	(M$_{\odot}$)	&	($10^{-12}$g cm$^{-3}$)	&		&		&	(cm$^2$ g$^{-1}$)	&	 (days)	&	(days)	&	\\
\hline\hline																						
PS1-12cil	&	$0.14_{-0.02}^{+0.03}$	&	$4.95_{-0.73}^{+0.76}$	&	$3.57_{-0.06}^{+0.07}$	&	$4.66_{-0.70}^{+0.76}$	&	$21.47_{-8.16}^{+7.58}$	&	 $0.43_{-0.11}^{+0.11}$	&	$0.11_{-0.01}^{+0.02}$	&	$0.01_{-0.01}^{+0.02}$	&	$-27.95_{-0.21}^{+0.22}$	&	$29.16_{-0.42}^{+0.34}$	&	$60.45/12$\\
2012aa	&	$0.20_{-0.00}^{+0.00}$	&	$5.89_{-0.12}^{+0.12}$	&	$1.63_{-0.01}^{+0.01}$	&	$1.47_{-0.09}^{+0.15}$	&	$2.14_{-0.85}^{+2.24}$	&	 $0.51_{-0.16}^{+0.14}$	&	$0.11_{-0.01}^{+0.02}$	&	$0.01_{-0.01}^{+0.01}$	&	$-46.46_{-0.47}^{+0.45}$	&	$24.72_{-5.22}^{+0.82}$	&	$92.90/22$\\
\hline\hline																		
\end{tabular}}
\end{center}
\end{table*}

\begin{table*}[tbp]
\tabcolsep=2pt
\caption{Parameters of the magnetar plus CSI model. The uncertainties are 1$\sigma$.}
\label{tab:mag+CSI}
\begin{center}
{\scriptsize
\begin{tabular}{cccccccccccccccc}
\hline\hline																		
	&	$\kappa$	&	$M_{\mathrm{ej}}$	&	$P_0$	&	$B_p$	&	$M_{\mathrm{CSM}}$	&	$\rho_{\mathrm{CSM,in}}$	&	$\epsilon$	&	$x_{\mathrm{0}}$	 &	$\kappa_{\gamma,\mathrm{mag}}$	&	$t_\mathrm{expl}$	&	$t_\mathrm{CSI}$	&	$\chi^2/\mathrm{dof}$\\
	&	(cm$^2$ g$^{-1}$)	&	(M$_{\odot}$)	&	(ms)	&	($10^{14}$~G)	&	(M$_{\odot}$)	&	($10^{-12}$g cm$^{-3}$)	&		&		&	(cm$^2$ g$^{-1}$)	&	(days)	&	(days)	&	\\
\hline\hline																								
PS1-12cil	&	$0.08_{-0.02}^{+0.04}$	&	$10.59_{-3.37}^{+3.25}$	&	$6.52_{-0.10}^{+0.09}$	&	$5.09_{-0.13}^{+0.13}$	&	$9.64_{-3.04}^{+2.97}$	&	 $56.09_{-21.01}^{+22.04}$	&	$0.18_{-0.05}^{+0.11}$	&	$0.12_{-0.01}^{+0.01}$	&	$2.72_{-0.40}^{+16.94}$	&	$-26.24_{-0.21}^{+0.20}$	&	 $28.64_{-0.21}^{+0.30}$	&	$54.39/11$\\
2012aa	&	$0.19_{-0.01}^{+0.01}$	&	$10.28_{-0.61}^{+0.72}$	&	$8.99_{-0.14}^{+0.11}$	&	$4.68_{-0.29}^{+0.40}$	&	$1.93_{-0.12}^{+0.15}$	&	 $4.24_{-2.42}^{+1.91}$	&	$0.29_{-0.10}^{+0.07}$	&	$0.12_{-0.01}^{+0.03}$	&	$0.01_{-0.01}^{+0.01}$	&	$-49.91_{-0.62}^{+0.83}$	&	 $25.32_{-0.35}^{+0.31}$	&	$36.77/21$\\
\hline\hline																		
\end{tabular}}
\end{center}
\end{table*}

\begin{table*}[tbp]
\caption{Parameters of the double CSI model. The uncertainties are 1$\sigma$.}
\label{tab:multiCSI}
\begin{center}
{\scriptsize
\begin{tabular}{cccccccccccccccc}
\hline\hline																		
	&	$i$th	&	$s$\tablenotemark{a}	&	$\kappa$	&	$M_{\mathrm{ej},i}$\tablenotemark{b}	&	$M_{\mathrm{CSM},i}$\tablenotemark{c}	&	 $R_{\mathrm{CSM,in},i}$\tablenotemark{d}	&	$\epsilon_i$\tablenotemark{e}	&	$\rho_{\mathrm{CSM,in},i}$\tablenotemark{f}	&	$x_{\mathrm{0}}$	&	 $t_{\mathrm{CSI},i}$\tablenotemark{g}	&	$\chi^2/\mathrm{dof}$\\		
	&	Interaction	&		&	(cm$^2$ g$^{-1}$)	&	(M$_{\odot}$)	&	(M$_{\odot}$)	&	($10^{14}$cm)	&		&	($10^{-12}$g cm$^{-3}$)	&		&	 (days)	&	\\		
\hline\hline
\multirow{4}*{PS1-12cil}	&	1	&	0	&	\multirow{2}*{$0.16_{-0.02}^{+0.02}$}	&	$13.18_{-2.07}^{+2.56}$	&	$7.13_{-1.15}^{+1.39}$	&	 $11.27_{-6.47}^{+9.31}$	&	$0.56_{-0.10}^{+0.14}$	&	$7.04_{-2.29}^{+3.03}$	&	\multirow{2}*{$0.14_{-0.01}^{+0.01}$}	&	$-24.46_{-0.27}^{+0.30}$	&	 \multirow{2}*{$21.08/10$}\\
	&	2	&	0	&		&	$20.30$	&	$6.26_{-0.97}^{+0.93}$	&	$80.53$	&	$0.14_{-0.02}^{+0.04}$	&	$1.60_{-0.37}^{+0.25}$	&		&	 $28.98_{-0.30}^{+0.38}$	&	\\
	&	1	&	2	&	\multirow{2}*{$0.15_{-0.01}^{+0.02}$}	&	$25.64_{-3.12}^{+2.51}$	&	$17.19_{-1.88}^{+1.59}$	&	$8.36_{-1.60}^{+1.71}$	&	 $0.69_{-0.11}^{+0.12}$	&	$7.23_{-2.17}^{+3.74}$	&	\multirow{2}*{$0.11_{-0.01}^{+0.01}$}	&	$-23.86_{-0.20}^{+0.20}$	&	\multirow{2}*{$31.48/10$}\\
	&	2	&	0	&		&	$42.84$	&	$4.07_{-0.51}^{+0.65}$	&	$77.32$	&	$0.27_{-0.04}^{+0.05}$	&	$0.72_{-0.24}^{+0.33}$	&		&	 $29.35_{-0.56}^{+0.35}$	&	\\
\hline																						
\multirow{4}*{2012aa}	&	1	&	0	&	\multirow{2}*{$0.19_{-0.01}^{+0.01}$}	&	$15.60_{-0.97}^{+0.93}$	&	$17.77_{-0.96}^{+0.95}$	&	 $8.15_{-3.99}^{+4.22}$	&	$0.30_{-0.02}^{+0.02}$	&	$18.19_{-1.71}^{+1.12}$	&	\multirow{2}*{$0.14_{-0.00}^{+0.00}$}	&	$-45.74_{-0.65}^{+0.61}$	&	 \multirow{2}*{$18.61/20$}\\
	&	2	&	0	&		&	$33.36$	&	$2.16_{-0.19}^{+0.20}$	&	$77.91$	&	$0.11_{-0.01}^{+0.01}$	&	$1.21_{-0.29}^{+0.28}$	&		&	 $25.07_{-0.50}^{+0.39}$	&	\\
	&	1	&	2	&	\multirow{2}*{$0.19_{-0.01}^{+0.00}$}	&	$19.46_{-1.25}^{+1.70}$	&	$19.41_{-0.65}^{+0.37}$	&	$19.79_{-2.34}^{+1.70}$	&	 $0.74_{-0.15}^{+0.09}$	&	$1.33_{-0.26}^{+0.36}$	&	\multirow{2}*{$0.11_{-0.01}^{+0.01}$}	&	$-43.28_{-0.60}^{+0.60}$	&	\multirow{2}*{$44.23/20$}\\
	&	2	&	0	&		&	$38.87$	&	$2.36_{-0.17}^{+0.15}$	&	$81.39$	&	$0.23_{-0.05}^{+0.03}$	&	$1.13_{-0.35}^{+0.31}$	&		&	 $19.26_{-0.10}^{+0.10}$	&	\\
\hline\hline																		
\end{tabular}}
\end{center}
\par
\tablenotetext{a}{$s$ is the density index of the $i$th CSM ($\rho_{\rm CSM} \propto R_{\rm CSM}^{-s}$).}
\tablenotetext{b}{$M_{\mathrm{ej},i}$ is the initial ejecta mass of the $i$th interaction, and $M_{\mathrm{ej},2}$ is calculated by $M_{\mathrm{ej},2} = M_{\mathrm{ej},1} +M_{\mathrm{CSM},1}$.}
\tablenotetext{c}{$M_{\mathrm{CSM},i}$ is the mass of the $i$th collided CSM.}
\tablenotetext{d}{$R_{\mathrm{CSM,in},i}$ is the initial radius of the $i$th collided CSM, and $R_{\mathrm{CSM,in},2}$ can be calculated by $R_{\mathrm{CSM,in},2} = R_{\mathrm{CSM,in},1} + v_{\mathrm SN}(t_{{\mathrm{CSI}},2} - t_{{\mathrm{CSI}},1})$.}
\tablenotetext{e}{$\epsilon_i$ is the efficiency of the kinetic energy convert to radiation of the $i$th interaction.}
\tablenotetext{f}{$\rho_{\mathrm{CSM,in},i}$ is the density of the CSM at $R_{\mathrm{CSM,in},i}$.}
\tablenotetext{g}{$t_{\mathrm{CSI},i}$ is the time of the $i$th interaction between the ejecta and the $i$th CSM.}
\end{table*}

\clearpage

\clearpage

\clearpage

\end{document}